\documentstyle[prl,aps,multicol,mypsfig2]{revtex}

\def\be{\begin{equation}}
\def\ee{\end{equation}}
\def\bea{\begin{eqnarray}}
\def\eea{\end{eqnarray}}

\def\>{\rangle}
\def\<{\langle}

\def\lb{\left[}
\def\rb{\right]}
\def\lp{\left(}
\def\rp{\right)}
\def\lbL{\lb\rule{0pt}{2.4ex}}

\newcommand{\mattwoc}[4]{\left[
	\begin{array}{cc}{#1}&{#2}\\{#3}&{#4}\end{array}\right]}
\newcommand{\vectwoc}[2]{\left[
	\begin{array}{c}{#1}\\{#2}\end{array}\right]}

\def\<{\langle}
\def\>{\rangle}

\begin{document}

\onecolumn

\title{Quantum algorithm for distributed clock synchronization}

\author{%
	Isaac L. Chuang\thanks{Electronic address: 
				ichuang@almaden.ibm.com
			\hfill 
			}
}

\address{\vspace*{1.2ex}
		IBM Almaden Research Center, 650 Harry Road \\
		San Jose, CA 95120 }
\maketitle

\vspace*{3ex}
\begin{abstract}
The clock synchronization problem is to determine the time difference
$\Delta$ between two spatially separated clocks.  When message delivery
times between the two clocks is uncertain, $O(2^{2n})$ classical messages must
be exchanged between the clocks to determine $n$ digits of $\Delta$.  On the
other hand, as we show, there exists a quantum algorithm to obtain $n$
digits of $\Delta$ while communicating only $O(n)$ quantum messages.
\vspace*{3ex}
\end{abstract}

\begin{multicols}{2}[]


Clock synchronization is an important problem with many practical and
scientific applications\cite{Simons88a,Lewandowski99a}.  Accurate
timekeeping is at the heart of many modern technologies, including
navigation, (the global positioning system), electric power generation
(synchronization of generators feeding into national power grids), and
telecommunication (synchronous data transfers, financial transactions).
Scientifically, clock synchronization is key to projects such as long
baseline interferometry (distributed radio telescopes), gravitational wave
observation (LIGO), tests of the general theory of relativity, and
distributed computation.

The basic problem is easily formulated: determine the time difference
$\Delta$ between two spatially separated clocks, using the minimum
communication resources.  Generally, the accuracy to which $\Delta$ can be
determined is a function of the clock frequency stability, and the
uncertainty in the delivery times for messages sent between the two clocks.
Given the stability of present clocks, and assuming realistic bounded
uncertainties in the delivery times (e.g. satellite to ground transmission
delays), protocols have been developed which presently allow determination
of $\Delta$ to accuracies better than $100$ ns (even for clock separations 
greater than $8000$ km); it is also predicted that accuracies of $100$ ps
should be achievable in the near future.

However, these protocols fail if the message delivery time is too uncertain,
because they rely upon the law of large numbers to achieve a constant
average delivery time (thus, also requiring $O(2^{2n})$ messages to obtain $n$
digits of $\Delta$).  If the required averaging time is longer than the
stability time of the local clocks, then these protocols must be replaced.
A simple, different, protocol, which succeeds independent of the delivery
time, is to just send a clock which keeps track of the delivery time.  For
example, if Alice mails Bob a wristwatch synchronized to her clock, then
when Bob receives it he can clearly calculate the $\Delta$ for their two
clocks from the difference between his time and that given by the
wristwatch.

This wristwatch protocol is generally impractical, but it suggests another
scheme which is intriguing.  A quantum bit (qubit) behaves naturally much
like a small clock.  For example, a nuclear spin in a magnetic field
precesses at a frequency given by its gyromagnetic ratio times the magnetic
field strength.  And an optical qubit, represented by the the presence or
absence of a single photon in a given mode, oscillates at the frequency of
the electromagnetic carrier.  The relative phase between the $|0\>$ and
$|1\>$ states of a qubit thus keeps time, much like a clock, and ticks away
during transit.  Unlike a classical clock, however, this phase information
is lost after measurement, since projection causes the qubit to collapse
onto either $|0\>$ or $|1\>$, so repeated measurements and many qubits are
necessary to determine $\Delta$.  On the other hand, with present technology
it is practical to communicate qubits over long distances through
fibers\cite{Hughes95a,Muller96a}, and even in free space\cite{Buttler98a}.

Here, we study this ``ticking qubit'' protocol for clock synchronization, and
establish an upper bound on the number of qubits which must be transmitted
in order to determine $\Delta$ to a given accuracy.  Surprisingly, we find
that only $O(n)$ qubits are needed to obtain $n$ bits of $\Delta$, if
we have the freedom of sending qubits which tick at different frequencies.
We begin by describing a formal model for this protocol, then the algorithm
is presented.  Various generalizations and limitations are discussed in the
conclusion.

\label{sec:protocol}

Let $t^{a}$ and $t^{b}$ be the local times on Alice and Bob's respective
clocks.  We may assume for now, for the sake of simplicity, that their
clocks operate at exactly the same frequency and are perfectly stable.
Their goal is to determine the difference $\Delta = t^b - t^a$, which is
initially unknown to either of them.  
The goal can be accomplished using the following primitive:

\parbox{3.30in}{
~\\
~\\
\centerline{{\bf Protocol: Ticking qubit handshake $\mathbf
TQH(\omega, |\psi\>)$ }} 
\begin{itemize}
\renewcommand{\makelabel}[1]{#1}
\setlength{\itemsep}{-1pt}

\item [\bf 1:] At time $t^a_1$, Alice sends $(t^a_1, |\psi\>, \omega)$ to
	Bob.  $\omega$ specifies the tick rate of the qubit $|\psi\>$.

\item [\bf 2:] Bob receives $(t^a_1, e^{i\omega t_{12} Z} |\psi\>, \omega)$
	at time $t^b_2$, where $t_{12}$ is the time the qubit spent in
	transit.

\item [\bf 3:] Bob applies the operation $C_{12} = X e^{-i\omega(t^b_2 -
	t^a_1)Z}$ to the qubit, obtaining $X e^{-i\omega Z \Delta} |\psi\>$.

\item [\bf 4:] At time $t^b_4$, Bob sends $(t^b_4, X e^{-i\omega Z \Delta}
	|\psi\>, \omega)$ to Alice.

\item [\bf 5:] Alice receives $(t^b_4, e^{i\omega t_{45} Z} X e^{-i\omega
	\Delta Z}|\psi\>, \omega)$ at time $t^a_5$, where $t_{45}$ is the
	time the qubit spent in transit.

\item [\bf 6:] Alice applies the operation $C_{45} = X e^{-i\omega(t^a_5 -
	t^b_4)Z}$ to the qubit, obtaining $e^{-2i\omega Z \Delta} |\psi\>$.

\end{itemize}%
}

We use notation for quantum states and their transforms that is standard in
the quantum computation and quantum information community; for an excellent
review, see~\cite{Bennett2000a}.  This can be summarized as follows.
$|\psi\>$ is the state of a qubit, which can be expressed as a two-component
unit vector
\be
	|\psi\> = c_0 |0\> + c_1 |1\> = \vectwoc{c_0}{c_1}
\,,
\ee
where $c_0$ and $c_1$ are complex numbers satisfying
$|c_0|^2+|c_1|^2=1$.  When measured, a $0$ results, projecting the
qubit into the state $|0\>$ with probability $|c_0|^2$; the
corresponding happens for $1$..  Operations on qubits are unitary
transformations $U$ which are matrices that satisfy $U^\dagger U = I$,
$U^\dagger$ being the complex-conjugate transpose of $U$ and $I$ the
identity matrix.  For single qubits, any $2$$\times$$2$ unitary
transform may be written a rotation operator,
\be
	e^{i\alpha + i\theta({n}_x X + {n}_y Y+ {n}_z Z)/2}
\,,
\ee
where $\alpha$ specifies a (usually irrelevant) global phase, $X$, $Y$,
and $Z$ are the usual Pauli matrices,
\be
	X =
	\mattwoc{0}{1}{1}{0}
~~~~~~
	Y =
	\mattwoc{0}{-i}{i}{0}
~~~~~~
	Z =
	\mattwoc{1}{0}{0}{-1}
\,,
\ee
$\hat{n} = [n_x,n_y,n_z]^T$ is a unit real-component vector, and $\theta$ is
the rotation angle.  Note that $X$, $Y$, and $Z$ themselves are valid
unitary operators; simple operations such as these are often called quantum
logic gates, and cascading them gives a quantum circuit.

The six stages of the $\mathbf TQH(\omega, |\psi\>)$ protocol work in
the following ways.  $\omega$ and $|\psi\>$ are inputs, as described
in Step~{\bf 1}; we will show below how they can be set usefully.  Step~{\bf
2} follows from the the time-evolution of the qubit during transit.  A
quantum state $|\psi\>$ evolves in time according to the Schr\"odinger
equation
\be
	-i\hbar \frac{d}{dt} |\psi\> = {\cal H} |\psi\>
\,,
\ee
where $\cal H$ is the (time-independent) Hamiltonian describing the
physical configuration of the system.  For example, a spin-$1/2$
particle such as an electron or proton in a magnetic field $B$ has the
Hamiltonian ${\cal H} = \hbar\omega Z$, where $\hbar\omega$ is the
energy difference between the state of the spin aligned and
anti-aligned with $B$.  Many other quantum systems, such as a single
photon propagating in space, can have a Hamiltonian of this
mathematical form.  Plugging this $\cal H$ into the solution to the
Schr\"odinger equation,
\be
	|\psi(t)\> = e^{i {\cal H} t/\hbar} |\psi(t=0)\>
\,,
\ee
gives $e^{i\omega t_{12} Z} |\psi\>$ after the elapsed time $t_{12}$.
Step~{\bf 3} is true because $t^b = t^a + \Delta$, and $t_{12} = t^b_2
- t^b_1 = t^b_2 - t^a_1 - \Delta$, so $C_{12} e^{i\omega t_{12} Z}
|\psi\> = e^{-i\omega Z \Delta} |\psi\>$.  During the time Bob has the
qubit, we assume he's turned off its evolution, so that although
$t^b_4$ may not equal $t^b_2$, the qubit does not experience any
relative phase shift during that time interval.  Step~{\bf 6} follows
because $t_{45} = t^a_5 - t^a_4 = t^a_5 - t^b_4 + \Delta$, and $X
e^{i\theta Z} X = e^{-i\theta Z}$.  Summarizing, the net effect of this
protocol is to allow Alice to transform a qubit $|\psi\>$ into
$e^{-2i\omega Z \Delta} |\psi\>$.

\label{sec:algorithm}

A simple, but inefficient, algorithm which allows Alice to determine
$\Delta$ uses repeated execution of the ticking qubit handshake.  She
prepares $|\psi\> = (|0\>+|1\>)/\sqrt{2}$, executes $\mathbf
TQH(\omega, |\psi\>)$, and obtains
\be
	|\psi'\> = \frac{e^{-2i\omega \Delta}}{\sqrt{2}} |0\>  
		+ \frac{e^{2i\omega \Delta}}{\sqrt{2}} |1\>
\,.
\ee
She then applies a Hadamard transformation
\be
	H = \frac{1}{\sqrt{2}}\mattwoc{1}{1}{1}{-1}
\ee
to $|\psi'\>$, getting
\be
	\frac{e^{-2i\omega \Delta} + e^{2i\omega \Delta}}{{2}} |0\>
	+ \frac{e^{-2i\omega \Delta} - e^{2i\omega \Delta}}{{2}} |1\>
\,,
\ee
such that when she measures the state, a $0$ results with probability
$\cos^2(2\omega\Delta)$.  By the law of large numbers, with high
probability, $2^{2n}$ repetitions of this procedure allows Alice to estimate
$n$ bits of $\cos^2(2\omega\Delta)$, and thus, of $\omega\Delta$.  If bounds
on the size of $\Delta$ are known in advance, $\omega$ can be chosen wisely
to allow $\Delta$ to be determined; otherwise, a few iterations of this
procedure with different $\omega$ suffice to initially bound $\Delta$.

This repetition procedure is inefficient because it requires an a number of
repetitions exponential in the number of desired digits.  It is essentially
a classical technique, and is very similar in structure to the usual
procedure employed in metrology, Ramsey interferometry\cite{Bollinger96a}.
The preparation of $|\psi\>$ can be accomplished by applying a Hadamard
transformation to $|0\>$; this corresponds to the first pulse in the Ramsey
scheme.  Note, incidentally, that in practice, Hadamard transforms can be
replaced with simple $\pi/2$ pulses (operations such as $e^{i\pi Y/4}$),
although their introduction here is convenient.  The $\mathbf TQH$ step
corresponds to the free evolution period.  And the final Hadamard is the
second pulse in Ramsey sequence.  It is thus not surprising that repetition
is inefficient, since it has the same resource requirements as in Ramsey
interferometry.

A much better algorithm, which allows Alice to determine $n$ bits of
$\Delta$ using only $O(n)$ repetitions of the ticking qubit handshake, is
the following.  Alice starts with $m+1$ qubits initialized to $|\phi_0\> =
|0\>|0\>$, where the decimal base label on the left denotes the $m$ qubit
state, and the label on the right, the extra ancillary single qubit.  She
then applies $m$ Hadamard gates to the $m$ qubits, obtaining
\be
	|\phi_1\> = \frac{1}{\sqrt{2^m}} \sum_{k=0}^{2^m-1} |k\> |0\>
\,.
\ee
The first register is now in an equal superposition over all possible states
of the $m$ qubits.  Next, Alice applies a unitary operation $T$ which does
\be
	T|k\>|0\> = |k\> e^{2 \pi i k \omega \Delta}|0\>
\,.
\label{eq:ttransform}
\ee
This is a nontrivial operation, but assume for now that this is possible and
below, we'll show how this is accomplished.  Applying $T$ to $|\phi_1\>$
gives
\be
	|\phi_2\> = \frac{1}{\sqrt{2^m}} \sum_{k=0}^{2^m-1} 
			e^{2 \pi i k \omega \Delta}
			|k\> |0\>
\,.
\ee
Next, Alice applies an inverse {\em quantum Fourier transform} $F^{-1}$,
which does
\be
	F^{-1} |k\> = \frac{1}{\sqrt{2^m}} \sum_{j=0}^{2^m-1}
			e^{-2\pi i j k/2^m} |j\>
\,.
\ee
This operation requires only $O(n^2)$ elementary one and two-qubit
gates\cite{Coppersmith94a} (in contrast to the classical fast Fourier
transform, which requires $O(n2^n)$ gates to transform an $n$ element
vector).  This produces the state (dropping the final $|0\>$, which is now
unimportant)
\be
	|\phi_3\> = \frac{1}{2^m} \sum_{k=0}^{2^m-1} \sum_{j=0}^{2^m-1}
		e^{2\pi i k (\omega \Delta - j/2^m)} |j\>
	= \sum_{j=0}^{2^m-1} c_j |j\>
\,.
\ee
$|c_j|^2$ is clearly peaked around $j=2^m \omega\Delta$.  If $2^m
\omega\Delta$ is an integer, then this equality holds, $|c_j|^2 =
\delta_{j,\omega\Delta}$, and measuring the first $m$ qubits gives
$\omega\Delta$ exactly.  Otherwise, it can be shown that if $m = n +
\lceil{\log(2+1/2\epsilon)}\rceil$, then measuring the $m$ qubits gives
$\omega\Delta$ to $n$ bits of accuracy, with probability of success at least
$1-\epsilon$\cite{Cleve98a,Nielsen00a}.

What we have used in this algorithm is the well-known ability of quantum
computation to efficiently determine the eigenvalue of a unitary operator,
for a given eigenstate, using a routine known as {\em quantum phase
estimation}\cite{Cleve98a,Nielsen00a}.  It is possible to use this
subroutine in the present application, clock synchronization, because there
can be an efficient implementation of the operator $T$.

\begin{figure}[htbp] 
\narrowtext
\begin{center}
\mbox{\psfig{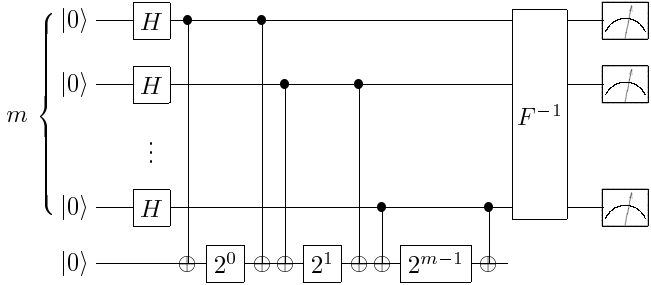}~~~~}
\end{center}
\caption{ Quantum circuit describing the exponentially fast clock
synchronization algorithm.  The boxed $2^{\ell}$ gates represent calls to
${\mathbf TQH}(-\pi 2^{\ell-1} \omega, |\psi\>)$; otherwise, the notation
used is standard\protect\cite{Barenco95a}.  Time goes from left to right,
and meters denote projective measurement.  }
\label{fig:circuit}
\end{figure}

Alice can implement $T$ using $m$ calls to $\mathbf TQH$.
One call is made for each of the $m$ qubits, so we can understand how this
works by considering what happens for the $\ell{\rm th}$ qubit.  Let $c_0
|0\> + c_1|1\>$ be the state of this qubit, so that we start with the
two-qubit state
\be
	(c_0 |0\> + c_1|1\>)|0\>
\,.
\ee
Now apply a controlled-{\sc not} gate\cite{Barenco95a}, whose transform is
described by the unitary matrix
\be
	\left[
	\begin{array}{cccc}
		1 & 0 & 0 & 0 \\
		0 & 1 & 0 & 0 \\
		0 & 0 & 0 & 1 \\
		0 & 0 & 1 & 0 
	\end{array}\right]
\,,
\ee
with the control qubit being the first one.  The result is
\be
	c_0 |00\> + c_1 |11\>
\,.
\ee
Note how the two qubits are now {\em entangled} --- this is partially
reflected by the fact that if a measurement were performed at this moment on
one qubit, the result would completely determine the state of the other
qubit.  Let $|\psi\>$ represent the state of the second qubit; Alice sends
this to Bob, performing ${\mathbf TQH}(-\pi 2^{\ell-1} \omega, |\psi\>)$,  Upon
completion of that procedure, she is left with the state
\be
	c_0 e^{2^\ell \pi i\omega \Delta/2} |00\> 
	+ c_1 e^{-2^\ell \pi i\omega \Delta/2} |11\> 
\,.
\ee
She then performs a second controlled-{\sc not} gate, again with the first
qubit as the control, obtaining
\bea
	\lbL{ c_0 e^{2^\ell \pi i\omega \Delta/2} |0\> 
	+ c_1 e^{-2^\ell\pi i\omega \Delta/2} |1\>  }\rb |0\>
\nonumber\\
~~	= e^{2^\ell \pi i \omega \Delta}
	  \sum_{k_\ell}  c_{k_\ell} |k_\ell\>
		e^{2^\ell\pi i k_\ell \omega \Delta}
		|0\>
\,.
\eea
The $e^{2^\ell i \omega \Delta}$ global phase is unobservable, and thus
irrelevant to the present calculation and can be dropped.  The overall
operation $T_\ell$ accomplished on this $\ell{\rm th}$ qubit can thus be
expressed as
\be
	T_\ell |k_\ell\> |0\> = |k_\ell \> 
		e^{2^\ell \pi i k_\ell \omega \Delta} |0\>
\,,
\ee
where $|k_\ell\>$ represents the $\ell{\rm th}$ qubit.  Now, the overall
state $|k\>$ of the $m$ qubits can be written as $|k\> =
|k_0\>|k_1\>\cdots|k_{m-1}\>$, so applying $T = T_0 T_1\cdots T_{m-1}$ gives
\bea
	T|k\>|0\> 
	= 
	   \lbL{ T_0 |k_0\>\, T_1|k_1\> \,
		\cdots
		T_{m-1}|k_{m-1}\> }\rb |0\>
\nonumber \\
~~
	=
 	   |k\> e^{ 2\pi i \omega\Delta \lp{ \sum_\ell 2^\ell k_\ell }\rp } 
		|0\>
\,.
\eea
Since $\sum_\ell 2^\ell k_\ell = k$, this construction gives the desired
transformation, Eq.(\ref{eq:ttransform}).  Note that the $m$ calls to
$\mathbf TQH$ can be performed sequentially (as shown in
Figure~\ref{fig:circuit}), or, by using $m$ ancilla qubits initialized to
$|0\>$, in parallel, since the algorithm leaves them unchanged.

The main caveat to this result is that the tick rate of the qubit $|\psi\>$
sent to Bob must span an exponentially large range, from $\sim 1/2\Delta$ to
$\sim 2^m/\Delta$.  If the qubit transmitted in the ${\mathbf TQH}$ routine
is physically realized by a spin in a magnetic field, this means that there
must be ``dial settings'' for the magnetic field strength which span an
exponentially large range.  Similarly, if the qubit is represented by a
single photon, its carrier must span an exponentially large frequency range.
Most critically of all, the stability of the tick rate must be adequate;
fluctuations of the magnetic field or index of refraction should be
controlled to cause less than roughly a half-wavelength phase shift.

On the other hand, in principle, if it is possible to use a nonlinear
optical medium to transport photons between Alice and Bob, then collective
photon states whose effective wavelengths can be exponentially
short\cite{Jacobson95} could be used to represent qubits.  Furthermore, the
shortest wavelength required by the protocol corresponds to the inverse of
the accuracy to which $\Delta$ is desired; this means that optical
wavelengths roughly correspond to accuracies of fractions of femptoseconds.
Time transfer using ground to satellite laser links is under
development\cite{Samain98a}, and photons of other wavelengths, ranging from
kilometers to millimeters, are also experimentally feasible.  The quantum
Fourier transform used in the quantum clock synchronization algorithm is
also known to be relatively stable to perturbations\cite{Barenco96a}, and
the entire procedure can be further stabilized by using quantum error
correction techniques\cite{Steane96a,Preskill98b,Knill97a,Gottesman98a}.

The quantum algorithm we have described allows two clocks to be
synchronized, independent of the uncertainties in message transport time
between the clocks, so long as messages are delivered within than the local
stability time of the clocks.  In its simplest instance, $2^{2n}$ ``ticking
qubit'' communication steps are required to obtain the time difference
$\Delta$ to $O(n)$ bits of accuracy.  Aside from exponential time, this does
not require any demanding physical resources -- just the ability to
communicate qubits.  In the advanced form of the algorithm, only $n$
``ticking qubit'' communication steps are required to obtain $O(n)$ bits of
$\Delta$, but this procedure requires exponentially demanding physical
resources.  These results invite further consideration of the problem of
clock synchronization with the assistance of quantum resources.  For
example, it is straightforward to simplify the present protocols to use only
one-way communication and no distributed entanglement (these results will be
reported in detail elsewhere).  It may also be possible to utilize quantum
teleportation\cite{Bennett93a} in a nontrivial manner, but that must be done
carefully, since changing the physical form of the qubits usually changes
their tick rate; moreover, the two classical bits sent in the teleportation
do not tick, and this is not apparently compensated by having an EPR pair
around.

We thank Daniel Gottesman and David DiVincenzo for helpful discussions.
 

\begin{thebibliography}{10}

\bibitem{Simons88a}
B. Simons, J.~L. Welch, and N. Lynch, IBM Research Report {\bf 6505},
  (1988).

\bibitem{Lewandowski99a}
W. Lewandowski, J. Azoubib, and W.~J. Klepczynski, Proc. IEEE {\bf 87},  163
  (1999).

\bibitem{Hughes95a}
R.~J. Hughes {\it et~al.}, Contemp. Phys. {\bf 36},  149  (1995),
  quant-ph/9504002.

\bibitem{Muller96a}
A. Muller, H. Zbinden, and N. Gisin, Europhys. Lett. {\bf 33},  334  (1996).

\bibitem{Buttler98a}
W.~T. Buttler {\it et~al.}, Phys. Rev. A {\bf 57},  2379  (1998).

\bibitem{Bennett2000a}
C. Bennett and D.~P. DiVincenzo, Nature {\bf 404},  247  (2000).

\bibitem{Bollinger96a}
J.~J. Bollinger, W.~M. Itano, D.~J. Wineland, and D.~J. Heinzen, Phys. Rev. A
  {\bf 54},  R4649  (1996).

\bibitem{Coppersmith94a}
D. Coppersmith, IBM Research Report RC 19642  (1994).

\bibitem{Cleve98a}
R. Cleve, A. Ekert, C. Macciavello, and M. Mosca, Proc. Roy. Soc. A {\bf 454},
  339  (1998).

\bibitem{Nielsen00a}
M.~A. Nielsen and I.~L. Chuang, {\em Quantum computation and quantum
  information} (Cambridge University Press, Cambridge, UK, 2000).

\bibitem{Barenco95a}
A. Barenco {\it et~al.}, Physical Review A {\bf 52},  3457  (1995),
  quant-ph/9503016.

\bibitem{Jacobson95}
J. Jacobson, G. Bjork, I. Chuang, and Y. Yamamoto, Physical Review Letters {\bf
  74},  4835  (1995).

\bibitem{Samain98a}
E. Samain and P. Fridelance, Metrologia {\bf 35},  151  (1998).

\bibitem{Barenco96a}
A. Barenco, A. Ekert, K.~A. Suominen, and P. Torma, Phys. Rev. A {\bf 54},  139
   (1996).

\bibitem{Steane96a}
A.~M. Steane, Physical Review Letters {\bf 77},  793  (1996).

\bibitem{Preskill98b}
J. Preskill, Proc. Roy. Soc. A: Math., Phys. and Eng. {\bf 454},  385  (1998).

\bibitem{Knill97a}
E. Knill and R. Laflamme, Phys. Rev. A {\bf 55},  900  (1997),
  quant-ph/9604034.

\bibitem{Gottesman98a}
D. Gottesman, Physical Review A {\bf 57},  127  (1998), quant-ph/9702029.

\bibitem{Bennett93a}
C.~H. Bennett {\it et~al.}, Phys. Rev. Lett. {\bf 70},  1895  (1993).

\end{thebibliography}

\end{multicols}
\end{document}